\documentclass[12pt]{article}
\usepackage{graphicx}
\begin{document}

\title{Investigating the influence of different thermodynamic paths on the 
structural relaxation in a glass forming polymer melt}

\author{Christoph Bennemann\\
	Wolfgang Paul\\
	J\"org Baschnagel \\
	Kurt Binder\\
        Institut f\"ur Physik\\
        55099 Mainz\\
	Germany\\
       	}

\maketitle
\begin{abstract}
We present results from Molecular Dynamics simulations of the thermal glass transition
in a dense polymer melt. In previous work we compared the simulation data with the idealized
version of mode coupling theory (MCT) and found that the theory provides a good description of the
dynamics above the critical temperature. In order to investigate the influence of different thermodynamic
paths on the structural relaxation ($\alpha$-process), 
we performed simulations 
for three different pressures
and are thus able to give a sketch of the critical line of 
MCT in the pressure-temperature-plane [$(p,T)$-plane], 
where,
according to the idealised version of MCT, an ergodic-nonergodic transition should occur. 
Furthermore, by cooling our system along two different paths
(an isobar and an isochor), with the same
impact point on the critical line, 
we demonstrate that neither the critical temperature nor 
the exponent $\gamma$ depend on the chosen path. 

\medskip

\noindent submitted to {\em J. Phys.: Condens.\ Matter} 
\end{abstract}

\section{Introduction}

The understanding of the glass transition has been a longstanding problem of condensed
matter physics and materials science
\cite{jaeckle,jaeckle2,zallen,kenna}.
The term {\it glass transition} is used to describe a phenomenon, 
where a solidification of
the liquid without simultaneous crystallisation occurs.
In a narrow temperature region the viscosity of the material increases by
some 14 orders of magnitude, and, although the 
typical time scales of the system
change tremendously, no significant change in the structure is observed and the material remains
amorphous.

Experimental research in this area of scientific interest has been conducted for more 
than a hundred years \cite{jaeckle,kohlrausch}. Since 
then a large number of phenomenological theories were proposed,
such as the free volume theory
\cite{cohen1,cohen2,cohen3} or the Gibbs-Di-Marzio theory \cite{gibbs1,gibbs2,gibbs3,gibbs4},
which tried to explain the
characteristic features of the glass transition.
These theories developed models, mainly based on 
physical intuition, from which thermodynamic properties
and the temperature dependence of the viscosity, i.e.,
the Vogel-Fulcher law \cite{jaeckle}, could be derived. 
However, an explicit relationship between the model parameters
and the microscopic properties of the glass former remained
hard to establish, and a detailed description of the shape
of dynamic correlation functions was not attempted.

In recent years the mode coupling theory (MCT) was
successful in describing a broad range of features observed in experiments 
\cite{Li1,vanMegen,Li2,franosch,ctoelle}
and simulations \cite{baschnag1,baschnag2,kob2,kob3,kaemmerer2,kks,kks1,sfct,artikel}
(also see \cite{yipn} for an overview)
in a temperature region close to a critical temperature $T_c$, which in general lies above 
the empirically defined glass transition temperature $T_g$. The critical temperature is 
associated with a change in dynamics from a liquid-like to a solid-like behaviour without
any concomitant modification of the glass former's structure. 
In contrast to most earlier theories this theory starts from well known
microscopic dynamics and uses techniques already applied in the field
of critical dynamics to derive a set of dynamic equations for
the density correlation functions of the system \cite{sjlander,lgoetze,goetze1,goetze2}
(also see \cite{mschilling,kob1} for review articles). It
made novel predictions on the temporal behaviour of the density correlations, which could be
tested experimentally and generated new research in this area.

The ideal version of the theory states that 
at the critical temperature the glass former should show 
a transition from ergodicity to nonergodicity \cite{lgoetze}. However, in real systems this
transition is in general absent (an exception being (probably) some colloidal glasses \cite{vanMegen}), 
because additional relaxation processes, which are
not included in ideal MCT, assure that the system is able to relax even below the 
ideal transition temperature.
In contrast, the 
extended version of MCT \cite{lgoetze,goetze1,goetze2} 
does account for these additional processes 
in terms of activated hopping processes and
therefore predicts that the transition is indeed avoided. 
Since sufficiently
above the critical temperature the contribution 
of jump processes to the dynamics of the glass former is negligible, the
ideal version of MCT often describes experimental data in this temperature
regime very well. Only if one is sufficiently close to or even below the critical
temperature, one has to apply extended MCT \cite{Li1,baschnag2}. 
However, such temperatures are seldomly
accessible in computer simulations of polymer systems, 
at least if equilibrated melts are required.  
Thus, in the following section we will concentrate
on the predictions of ideal MCT.

Although MCT has been applied to experimental data numerous times,
many aspects of the theory still remain to be thoroughly investigated. Little is known
on the influence of external thermodynamic parameters on the 
transition temperature (although see \cite{ben86,lkohc,toelle,ptoelle}) or the $\alpha$-process.
Since in general higher pressure causes higher densities,
which in turn means that the movement of an individual particle is more hindered,
an increase in pressure results in an increase of the transition
temperature. This effect is well known for the glass 
transition temperature \cite{kenna}. Hence, it is possible to cause a glass transition
by sole increase of the pressure, which indeed has been observed in experiment \cite{ptoelle}.
In general the thermodynamic parameter space (for instance, the $(p,T)$-plane) 
should decompose into two areas, a fluid phase and an ideal glass phase, 
both separated by a critical line at which the transition to nonergodicity should
occur. 
Since MCT applies to systems in thermodynamic equilibrium, 
the position of the critical
line should be independent on the thermodynamic path chosen upon 
cooling. 
Furthermore, close to the critical line, the exponents of the theory,
which determine most of the quantitative behaviour of the glass former,
depend solely on the impact point of the thermodynamic path on the
critical line, granted one does not choose a too exotic path, e.g. 
one that runs almost parallel to the critical line.
Therefore two different thermodynamic paths, which have the same impact
point on the critical line, should not only yield the same critical
temperature, but the quantitative behaviour of the systems, described
in terms of MCT, should be the same along both of them.
To our knowledge, so far this prediction has never been 
verified. 

Thus, we decided to conduct a study of the influence of pressure
on the parameters of ideal MCT, which also should enable us to
give a sketch of the critical line in the $(p,T)$-plane. To this end, we chose a model for
a polymer melt, which has already been used in an earlier study of the
glass transition \cite{artikel}. Clearly, because of the connectivity of the monomers
along the chain, our model is by no means a simple liquid, and therefore
certainly not the kind of system ideal MCT was originally developed
for. On the other hand, polymers, according to Angell's 
classification scheme \cite{angell},
mostly belong to the fragile glass formers, to which MCT has been
applied successfully, and are extremely good glass formers, i.e. 
``supercooled'' melts in thermal equilibrium can be prepared very well. 
In previous work \cite{artikel} it was demonstrated 
that it is possible to equilibrate our model well in
the regime of the supercooled melt, a prerequisite to
apply MCT, and that the $\alpha$-relaxation behaviour is compatible with MCT. 
Since we were able
to simulate the system in the isochoric (NVT) as well as in the isobaric (NpT)
ensemble, we try to test the prediction of thermodynamic 
path independence in the present paper. To this end, we cooled our system along
an isochoric path, which shared the impact point on the critical line
with one of the isobaric paths.
As in our earlier work \cite{artikel}, we concentrate 
on the $\alpha$-relaxation behaviour in this study. An 
analysis of the $\beta$-relaxation can be found
in Ref.~\cite{supermct}.

The remainder of the paper is organised as follows. In section II we briefly
describe our model and the simulation technique. 
By performing simulations along different isobars 
we were able to investigate the influence of pressure on the 
critical temperature. The results of these simulations
will be discussed in section III.
In section IV we will perform a test of the 
thermodynamic path independence, and in section V conclusions will be drawn.

\section{Model and Simulation Technique}\label{secII}

For modelling the inter- and intramolecular forces we used
a bead-spring model derived from the one suggested by Kremer 
and Grest \cite{kremer} and also used in several recent simulations
 \cite{duenweg1,duenweg2}. However, we included here also the attractive
part of the Lennard-Jones potential, since previous work on
a lattice model for a glassy polymer melt \cite{Wolf1,Wolf2}
had shown that without such an attraction the model would produce a negative
thermal expansion coefficient. The model of Kremer et al. \cite{kremer,duenweg1,duenweg2} is close to an athermal model of polymer melts and hence does not
exhibit a glass transition driven by temperature at all.

As in our past simulations each chain consisted of 10 beads with mass $m$ set 
to unity. Although these chains are rather short, they already show the static
behaviour characteristic of long polymers in the melt (e.g. Gaussian statistics
for the end-to-end distance distribution, a Debeye scattering law for the
single chain structure factor, etc.). Note that each bond in this model would
correspond to $n\approx3-6$ covalent bonds along the backbone of a real chain,
if one were to map this coarse-grained model onto a real polymer. Between
all monomers there acted a truncated Lennard-Jones potential:

\begin{equation}
U_\mathrm{LJ}(r_{ij}) = \left\{ \begin{array}{r@{\quad:\quad}l}
4\epsilon \left[ \left(\frac{\sigma}{r_{ij}}\right)^{12}
- \left(\frac{\sigma}{r_{ij}}\right)^{6}\right] + C & r_{ij} < 2 \cdot 2^{\frac{1}{6}} \sigma\\
0 & r_{ij} \geq 2 \cdot 2^{\frac{1}{6}}\sigma 
\end{array} \right. ,
\end{equation}
where $C$ was a constant which guaranteed that the potential was continuous everywhere. Since
it was not our aim to simulate a specific polymer, we used Lennard-Jones
units, where $\epsilon$ and $\sigma$ are set to
unity. Note that this means that all quantities are dimensionless. 
In addition to the Lennard-Jones potential a FENE backbone potential was applied along
the chain:

\begin{equation}
U_\mathrm{F}(r_{ij})=-\frac{k}{2}R_{0}^2 
\ln\left[1-\bigg(\frac{r_{ij}}{R_0}\bigg)^2\right].
\end{equation}

\noindent
The parameters of the potential were taken as $k=30$ and $R_0=1.5$, guaranteeing a certain
stiffness of the bonds while avoiding high frequency modes and chain crossing. 
Furthermore, with these parameters we set the favoured bond length
to a value slightly smaller than the length favoured by the Lennard-Jones potential. Thus
we introduced two different incompatible length scales in our system, 
which prevents 
the emergence of long range order (i.e. crystal formation) 
at lower temperatures.

Unlike previous lattice models for the thermally driven glass transition of
polymers \cite{Wolf1,Wolf2}, the present model has a qualitatively 
reasonable equation of state
with a positive thermal expansion coefficient, and can easily be studied under
constant density or constant pressure. It allows to study motion and structure
from local scales (motions in the neighbour cage) upto large scales.

In order to keep the temperature fixed,
all simulations were performed using a Nos\'e-Hoover thermostat
 \cite{Nose1,Hoover1,Tolla}.
In this technique the  model system
is coupled to a heat bath, which represents an 
additional degree of freedom. 
To set the system to a desired pressure, the size of the 
simulation box was adjusted to yield the correct density at 
each temperature. The resulting configurations were used
as start configurations for runs in the canonical ensemble,
where the size of the simulation box was kept fixed. Only
during these canonical runs dynamic correlation functions
for the further analysis were calculated. A more thorough 
discussion of the simulation technique applied can be found 
elsewhere \cite{artikel,doktorarbeit}. Here we only emphasize that we have
carefully checked that the Nos{\'e}-Hoover thermostat does not lead to any
artefacts in the dynamics of the single chain correlators and local properties
that were studied here \cite{doktorarbeit}. Note also that our chain length
$N=10$ was short enough that our results are not affected at all by chain
entanglement effects.

Altogether we
performed simulations at more than $40$ different points in the
thermodynamic phase space. At each point $10$ independent configurations
were simulated, each consisting of $120$ polymer chains of $10$ monomers.
With this we were able to perform simulations along three 
isobars and one isochor. 
Table 1 shows which temperatures were simulated in which 
ensemble.

In order to equilibrate an individual system at lower temperatures
one had to simulate
for very long times ($> 10^6$ MD-steps). Generally, the equilibration
of the lowest temperature in a given ensemble lasted as long as 
the sum of all equilibration times at higher temperatures of the same ensemble.
Altogether the simulations consumed an equivalent of approximately 
$10$ CPU-years on a PentiumPro processor run at $180$ Mhz.

\section{Dynamical properties at different pressures}\label{sec3}

As already discussed in the opening paragraph the glass transition manifests itself
by a 
steep increase of the relaxation times by several orders of magnitude.
In order to extract these time scales from
 the simulation data we computed a number of dynamical
quantities, like the incoherent intermediate dynamic structure factor:
\begin{equation}
\phi_q^\mathrm{s}(t) = \left<\frac{1}{N}\sum_{i=1}^{N} \mathrm{e}^{
\mathrm{i} {\bf q} \cdot ({\bf r}_i(t)-{\bf r}_i(0))}\right> ,
\end{equation}
where $N$ stands for the total number of monomers in the melt. This function
measures the self correlation of the particle positions at different times,
and, by varying the wave-vector {\bf q}, at different length scales. 

Recently, orientational degrees of freedom and their relaxational behaviour 
have come into focus of theoretical research on the glass transition
\cite{oschilling,oschilling2}. { Results of molecular dynamic simulations for 
a fluid consisting of diatomic molecules \cite{kaemmerer2,kks,kks1} illustrated
that there can be significant differences between orientational 
and translational relaxation. Such differences are also
observed in experiments (see Ref.~\cite{lpdssl}, for instance).
Clearly, it should be interesting to check, whether we could find any differences
between orientational and translational relaxation in our model.
Hence, we also calculated
the orientational correlation of the end-to-end vector:
\begin{equation}
E_n(t)\equiv \left< L_n\left(\frac{{\bf e}(t) \cdot {\bf e}(0)}
{\parallel {\bf e}(t) \parallel \parallel {\bf e}(0) \parallel }\right) \right>, \;\; 
n=1,2,\ldots \; ,
\label{en}
\end{equation}
where $L_n$ stands for the $n$th Legendre-polynomial, ${\bf e}(t)$ is the end-to-end vector
of a polymer at time $t$, and $\parallel {\bf e}\parallel$ is
the length of the end-to-end vector at time $t$. The same 
formula can be applied to measure the dynamical correlation
of a bond vector ${\bf b}(t)$:
\begin{equation}
B_n(t)\equiv \left< L_n\left(\frac{{\bf b}(t) \cdot {\bf b}(0)}{\parallel {\bf b}(t) \parallel 
\parallel {\bf b}(0) \parallel }\right) \right>, \;\; n=1,2,\ldots \;.
\label{bn}
\end{equation}
Equations~(\ref{en}) and (\ref{bn}) characterize the reorientation
dynamics of the largest and of the smallest vectors along 
the backbone of a chain. In the analysis, we only calculated
the first and second polynomial, since these quantities
can be measured by dielectric relaxation and light
scattering, respectively.

With these three dynamical correlation functions we define the following correlation times:
\begin{equation}
\phi_q^\mathrm{s}(\tau_q) = 0.3 \: \: \: \quad E_n(\tau_{E_n}) = 0.3 \: \: \: \quad B_n(\tau_{B_n})=0.3 \; .
\label{taudef}
\end{equation}
We have computed a number of other related quantities as
well, such as the Rouse-modes of the system or the mean-square-displacements,
which are discussed in other publications \cite{artikel,supermct,rouse}.

\subsection{Behaviour of dynamical correlators}

As shown in Fig.~\ref{fig1} and Fig.~\ref{fig2}, the correlators decay 
in one step at high temperatures, while at lower temperatures
a two step process starts to emerge, 
which becomes the more pronounced, the lower the temperature.
The emergence of a plateau in the decay is related to the cage effect \cite{lgoetze}, 
where an individual monomer
is trapped by its surrounding particles. The average time a monomer needs to escape from
the cage
of its neighbours
increases with decreasing temperature, which explains the
extension of the plateau.
The presence of a two step relaxation, the so-called $\beta$-relaxation
(onto and off of the plateau) and $\alpha$-relaxation (off of
the plateau and long-time structural relaxation), 
is a common feature of glass formers, and is also predicted by MCT.

As can be seen from the plots, the qualitative behaviour is not affected by the
applied pressure, 
although at higher pressure the two step process starts to show up at 
higher temperatures already. Furthermore, while the height of the plateau 
depends on the specific correlator, it hardly varies with pressure.
The orientational correlators [first and second Legendre-polynomial, see 
Eqs.~(\ref{en}) and (\ref{bn})] exhibit a rather high plateau value
which is often close to unity (the plateau of the first Legendre-polynomial is 
always larger than that of the second). Therefore the
two step process is only visible on magnification. Clearly, the contribution
of the $\alpha$-process to the overall relaxation of a correlator depends 
on the quantity considered.

Another characteristic of glass forming liquids is that close to the critical temperature the
time temperature superposition principle should hold for the $\alpha$-relaxation. 
One therefore has to rescale a 
dynamical correlator by a suitably defined $\alpha$-relaxation time and to check whether
the curves fall on a master curve in the $\alpha$-regime.
As we reported in our earlier work for $p=1$ (and constant
volume) \cite{artikel}, 
this is indeed the case. Here, we additionally observe that our data
also obey a time-temperature-pressure superposition principle, i.e., in the $\alpha$-regime
data taken from different isobars collapse on a single master curve.
This is illustrated for a number of different dynamic correlators in Fig.~\ref{fig3}., where 
fourteen different curves are included in one plot.
Similar behaviour has been observed in experiments 
on orthoterphenyl \cite{toelle}, but, to the best of our 
knowledge, this is the first report from computer 
simulations for such a behavior.

\subsection{Behaviour of relaxation times}

Figure~\ref{fig1} and \ref{fig2} show that on lowering the temperature 
an increase of 
the relaxation times by several orders of magnitude takes place,
as expected for a glass forming liquid.
By means of asymptotic expansions the idealised MCT
derives a number of predictions concerning the behaviour of the liquid 
close to the critical temperature. One finds that sufficiently close to $T_c$
the increase of the $\alpha$-relaxation times can be described by the following
formula:
\begin{equation}
\tau = \tau^0 \left(T -T_c \right)^{-\gamma},
\label{tgamma}
\end{equation}
where 
$\tau^0$ is an amplitude which depends
on the specific relaxation time considered, and $\gamma$ is a parameter of the
theory which should be the same for all correlation times, if
the corresponding correlator couples to density fluctuations. Furthermore 
our analysis of the $\beta$-regime suggested that $\gamma$ should take
the value of $\gamma=2.09$ for the isobar $p=1.0$ \cite{supermct}.

At all pressures investigated, it is indeed possible to 
locate a temperature interval, where the increase of the 
relaxation times can be described by Eq.~(\ref{tgamma}).
When applying Eq.~(\ref{tgamma}), $\tau^0$, $T_c$, and
$\gamma$ were treated as adjustable parameters. 
Although it is not incompatible with MCT that rotational and
translational degrees of freedom freeze at different state points in the 
temperature-density plane, as was demonstrated in recent publications 
\cite{oschilling,oschilling2},
our analysis suggests that 
it is possible to find a critical temperature for all
isobaric paths, which is independent of the specific 
correlator and solely a function of the pressure considered.
The critical temperatures and densities we obtained are listed in Table \ref{tcres}.
Note that the error for the critical temperature for 
$p=1.0$ is smaller than for the
other pressures because a larger number of temperatures was 
simulated for this isobar. The pressure dependence of the 
critical temperature is depicted in Fig.~\ref{tcp}. As 
expected, the critical temperature increases with increasing 
pressure as also calculated for Lennard-Jones models in \cite{ben86}.
As one can also see from Table \ref{tcres} within the error bars the
quantity $\rho_c T_c^{-1/4}$ is a constant at the mode coupling
critical point as also found experimentally in e.g. \cite{ptoelle} and
in the simulation of soft sphere models \cite{bernu,roux,barrat} and for Lennard-Jones mixtures \cite{nauroth}. The value we found is within the error bars 
identical to the Lennard-Jones value in \cite{nauroth}.

Figure~\ref{tau} shows a double logarithmic plot of 
$\alpha$-relaxation times against $T-T_c$, using the critical 
temperatures of Eq.~(\ref{tcres}). For all pressures there
is a temperature interval, where the data points lie on a 
straight line in accord with Eq.~(\ref{tgamma}). Deviations
from the power-law behavior are visible both at small and
large distances from the critical point. The deviations for
large $T-T_c$ are expected because Eq.~(\ref{tgamma}) is
an asymptotic expansion which is only valid, if the
reduced distance to $T_c$, i.e., $(T-T_c)/T_c$, is small.
Upper bounds for the validity of Eq.~(\ref{tgamma}) are
approximately 0.7 ($p=0.5$), 1.2 ($p=1$), and 0.6 ($p=2$),
which is comparable to experiments \cite{Li1,Li2}
and other simulations \cite{kob2,kob3,kaemmerer2,kks,kks1}
However, these upper bounds strongly depend on the quantity
under consideration. Whereas deviations are very pronounced
for the smallest length scale ($q=9.5$), Eq.~(\ref{tgamma})
provides a good description at all, except perhaps at the
lowest temperatures for the end-to-end distance.

On the other hand, the deviations from the idealised power
law at low temperatures could be attributed to the 
ergodicity restoring jump processes mentioned above.
Close to the critical temperature these processes start to 
contribute significantly to the relaxation dynamics of the 
system, and therefore the actual relaxation times can be 
smaller than the predictions of idealised MCT. This
behavior has been discussed in experiments, e.g. \cite{Li1,Li2},
and simulations, e.g. \cite{baschnag2,kaemmerer2,kks}.
Therefore, in practical application of Eq.~(\ref{tgamma})
one faces the problem that its range of validity is
limited from below and above, and that it additionally
depends on the quantity under consideration.

Furthermore, Eq.~(\ref{tgamma}) implies that in the temperature regime, where the
idealised MCT is applicable, the ratio of two different $\alpha$-relaxation times
should be independent of temperature. As demonstrated in
Fig.~\ref{ratio}, this is not the case, even in the regime where the $\beta$-analysis
could be done, i.e., for $T-T_c\leq 0.07$ \cite{supermct}.
The ratio between different relaxation
times can change by almost a factor of two, and the effect is
stronger for $q=9.5$ than for $q=6.9$ (first minimum and maximum of the static structure 
factor, respectively). Note that we obtain the same result when applying a different definition
of the $\alpha$-relaxation time, which includes the nonergodicity parameter $f_q^\mathrm{sc}$,
i.e., $\phi_{q}^\mathrm{s}(\tau_q) = \mathrm{e}^{-1}f_q^\mathrm{sc}$.
However, it is not clear whether this finding is a strong contradiction to MCT, because
we have eliminated the dominant temperature dependence, given by Eq.~(\ref{tgamma}), when 
dividing two
relaxation times. Since we are close, but not very close to $T_c$, and Eq.~(\ref{tgamma}) is,
strictly speaking, only asymptotically valid, one could expect a smooth temperature
dependence of prefactors. Such a conclusion can also be drawn from Ref.~\cite{franosch}, in
which the MCT-equations for a model of a colloidal suspension are solved numerically and
compared with the asymptotic results. There, it is found that the ratio of two
relaxation times becomes constant only very close to the critical point, although Eq.~(\ref{tgamma})
is already observed for larger distance to the critical volume fraction (see Fig.~7 of 
Ref.~\cite{franosch}). Interestingly, Fig.~\ref{ratio} shows that the
ratio is not a monotonic function and exhibits a maximum approximately at the beginning of the 
temperature interval, in which we can apply ideal MCT to describe the $\alpha$-relaxation time. 
It seems as if at this temperature a change in the dynamics of the system occurs.

This problem is also reflected in Fig.~\ref{gammaq} which
shows the results for $\gamma$ when fitting Eq.~(\ref{tgamma}) 
to the $\alpha$-relaxation time of $\phi_q^\mathrm{s}(t)$ at $p=1$.
The critical temperature was kept fixed ($T_c=0.45$) in 
the fits, and the maximum possible number of temperatures was
taken into account to determine $\gamma$. It is interesting 
to note that the $\gamma$-values, determined from 
Fig.~\ref{tau} for the different pressures, agree with 
one another within the error margins so that the following
discussion is not specific for $p=1$.
Figure~\ref{gammaq} shows that the fit procedure yields a 
decrease of $\gamma$ with 
decreasing $q$, but the $\gamma$-values are distributed
around the result of the $\beta$-relaxation analysis,
$\gamma=2.09$ \cite{supermct}. 
Alternatively one can keep the exponent $\gamma=2.09$
constant  and adjust the critical 
temperature, $T_c$m, \cite{supermct}. Then
the critical temperatures for $q\geq 3$ coincide
within the error bars with the value obtained from the
$\beta$-analysis. However, the diffusion coefficient of a
chain yields a $T_c$, which is significantly lower. Physically, 
both types of analysis suggest that going from $T_c$ to higher temperatures
the melt has a stronger tendency to liquify
on short than on the large length scales. Such a behavior
is not unique to our polymer model, but It was also found in
other simulations \cite{kob2,kaemmerer2}.

\section{Test of the thermodynamic path independence}

In order to verify the prediction of the thermodynamic path independence, we estimated the 
density of the melt at the critical temperature ($T_c = 0.45$) of the isobar $p=1$. 
The density is $\rho = 1.042$. Then we performed a number of simulations in
the NVT-ensemble at the appropriate isochor (schematically, this is illustrated in
Fig.~\ref{tcp}), and again calculated various dynamic correlation
functions at the  simulated temperatures (see Table~\ref{tab1}).

The qualitative behaviour of the dynamical correlation functions along this isochor does
not differ from the behaviour observed at the various isobars, discussed in the preceding section. 
As can be seen in Fig.~\ref{fig8} and Fig.~\ref{fig9}, which show exemplarily 
the dynamic correlation of the end-to-end vector orientation,
we find again that at lower temperatures a two step relaxation occurs
(which cannot be seen on the scale of figure due to the large
plateau value), that the relaxation times
show a steep increase, and that at least for the lower temperatures the time temperature
superposition principle holds. 
This could have been expected,
since also earlier simulations of the model in the NVT ensemble
had shown such a behaviour \cite{artikel}.
It is interesting to note, however, that, compared to the 
appropriate isobar, the two step relaxation process can now 
be observed at higher temperatures already and that for the 
studied temperatures the $\alpha$-relaxation time is almost 
one order of magnitude larger. 

Once again it is possible to find a temperature region, where
the behaviour of the $\alpha$-relaxation times, extracted from  
the different dynamic correlation functions, can be described by Eq.~(\ref{tgamma}).
This is illustrated in Fig.~\ref{fig10}, where we show the temperature dependence of various
correlation times, as measured in the NVT-ensemble, plotted in such a way that the
applicability of the MCT-prediction is clearly demonstrated.
Qualitatively, we find the same features, as discussed 
before for the isobars. There are deviations from linearity
at large temperatures, the deviations are more pronounced
for the shortest length scales, but still the fits yield very 
similar values for the critical temperature, which can be
combined to:

\begin{equation}
T_c(\rho=1.042) = 0.445 \pm 0.010 \;.
\end{equation}

\noindent Within our error bars the value coincides with the critical temperature obtained for
the isobar $p=1.0$. For the NVT-simulation the error bar is larger, since we were
not able to equilibrate the melt as closely to the critical
temperature as it was possible in the NpT-simulation. Note that the lowest temperature
in Fig.~\ref{tau}b is $T=0.46$, whereas it is $T=0.5$ in Fig.~\ref{fig10}. This difference is caused
by the larger relaxation time in the NVT-ensemble (due to higher density/pressure)
at a specific temperature in comparison to the NpT-ensemble. Therefore the estimate
becomes less accurate, but we can still conclude
that the critical temperature of ideal MCT is
indeed independent of the thermodynamic path chosen.

Finally, we want to verify whether the exponent $\gamma$
is independent of the thermodynamic path as well. 
As already discussed in section III, the exponent $\gamma$ 
shows a pronounced dependence on the dynamic correlation 
function considered, if one works with the same critical 
temperature for all quantities and extends the fit interval
as much as possible. The same dependence is also found for
the isochor, but the results coincide within the error
bars with those of the isobars, as Fig.~\ref{fig10}
illustrates. Therefore, $\gamma$ is in fact independent of
the chosen path, which exemplifies the thermodynamic 
character of the critical point in mode-coupling theory.

\section{Conclusions}

In this paper we have presented results of a large scale 
molecular dynamics simulation for a supercooled polymer melt. 
Our model is a coarse-grained bead-spring model with 
nonlinear springs connecting monomers along a chain and 
Lennard-Jones interactions between all monomers. By including 
competing length scales in the model we
prevented the melt from crystallising at lower temperatures.

The present study concentrated on the influence of pressure
on the $\alpha$-relaxation behaviour of the melt. 
Upon cooling we see a steep increase of the $\alpha$-relaxation 
time, and all dynamic correlation
functions show a two-step relaxation. By comparing 
data from different isobars, we
found that our system does not only exhibit 
time-temperature-superposition above $T_c$, but
time-temperature-pressure-superposition as well.

For all pressures investigated it has been possible to 
locate a temperature interval, where the increase of the 
$\alpha$-relaxation times could be described by idealised 
MCT. Therefore we have been able to investigate the
dependence of the critical temperature of MCT on pressure 
and to give a sketch of the critical line in the 
$(p,T)$-plane. However, whereas the critical temperatures,
determined from different quantities probing both small and 
large length scales of the melt, coincide within the error
bars, the approach towards $T_c$, i.e., the exponent
$\gamma$, is very sensitive to the precise choice of 
$T_c$ in the fit, and depends on the quantity considered. 
When fixing $T_c$, we find that the $\alpha$-relaxation
times of $\phi_q^\mathrm{s}(t)$ for $q$-values distributed around 
the maximum of the structure factor are compatible with 
result of the $\beta$-analysis \cite{supermct}. Deviations
occur for much smaller and larger $q$-values. The
deviations at large $q$ can be explained by the sensitivity
of $\gamma$ on $T_c$, since fixing $\gamma$ at the value
of the $\beta$-analysis instead of $T_c$ yields estimates
for $T_c$ that are compatible with the result of the
$\beta$-analysis \cite{supermct}. However, such an 
alternative fit procedure does not remove the discrepancies 
found on the largest length scale. On these length scales, 
$\gamma$ is smaller than expected from the $\beta$-analysis.
Similar deviations are also
observed on smaller length scales, if the critical
point is approached very closely. They can be rationalized,
within the theoretical framework of MCT, by ergodicity
restoring processes, which compete with and finally dominate
over the cage effect treated by the idealised theory,
if $T \leq T_c$. To what extent the predictions of the
idealised theory are observable, therefore depends not only
on the quantity under consideration, which is also pointed
out in recent theoretical work \cite{franosch,fgm}, but also on the 
distance to the critical point. If one is too close, 
ergodicity restoring processes interfere, and if the temperature is
too large, one leaves the asymptotic regime, where the
formulas of the idealised MCT are expected to hold. 

By performing simulations along an isochor which had the same impact point
on the critical line as one of the isobars, we have been able to verify, 
that within the error margin the critical temperature of MCT is indeed 
independent of the thermodynamic path one chooses upon cooling. Furthermore, 
we have shown, that the exponent $\gamma$ does not depend on
the thermodynamic path either, within the caveats explained in the last paragraph. 

In summary, one can therefore say that
the idealised theory is a good starting point for a quantitative
description of the dynamics above $T_c$, and seems to capture
the essential physics, not only for simple liquids, but also for
polymer model. Reasons, why this could be the case, are further
discussed in Ref.~\cite{supermct}.

\section{Acknowledgements}
We would like to thank W. Kob, A. Latz and B. D\"unweg for stimulating discussions, and
also A. Kopf for supplying his MD-code. Support by the Sonderforschungsbereich
SFB 262, and a generous grant of computer time from the computing center
at the university of Mainz and the HLRZ J\"ulich, are gratefully acknowledged.

\clearpage

\begin{table}[t]
\caption[]{The table shows, at which temperatures and 
densities or pressures simu\-la\-tions were performed.}

\vspace{2mm}
\label{tab1}
\begin{tabular}{|l|l|}\hline
ensemble & simulated temperatures \\ \hline\hline
isochor ($\rho=1.042$) & $0.5$, $0.52$, $0.55$, $0.58$, $0.6$, $0.65$, $0.7$, $0.8$, $0.9$, $1.0$, $2.0$ \\ \hline
isobar  ($p=0.5$) & $0.45$, $0.48$, $0.5$, $0.52$, $0.55$, $0.6$, $0.7$, $1.0$ \\ \hline
isobar  ($p=1.0$) & $0.46$, $0.47$, $0.48$, $0.49$, $0.5$, $0.52$, $0.55$, $0.6$, $0.65$, $0.7$, $1.0$, $2.0$, $4.0$ \\ \hline
isobar  ($p=2.0$) & $0.52$, $0.55$, $0.57$, $0.6$, $0.7$, $0.8$, $0.9$, $1.0$, $2.0$ \\ \hline
\end{tabular}
\end{table}

\begin{table}
\caption[]{Critical temperatures and densities and soft sphere scaling variable
at the critical point}

\vspace{2mm}
\label{tcres}
\begin{tabular}{|l|l|l|l|}\hline
$p$ & $T_c$ & $\rho_c$ & $\rho_c T_c^{-1/4}$ \\\hline
$0.5$ & $0.425 \pm 0.010$ & $1.035 \pm 0.01$ & $1.28 \pm 0.02$ \\\hline
$1.0$ & $0.450 \pm 0.005$ & $1.042 \pm 0.01$ & $1.27 \pm 0.02$ \\\hline
$2.0$ & $0.490 \pm 0.010$ & $1.054 \pm 0.01$ & $1.26 \pm 0.02$ \\\hline
\end{tabular}
\end{table}

\clearpage

{\bf\Large Figure Captions}\\
\begin{figure}[h]
\centering
\includegraphics[width=110mm,height=90mm]{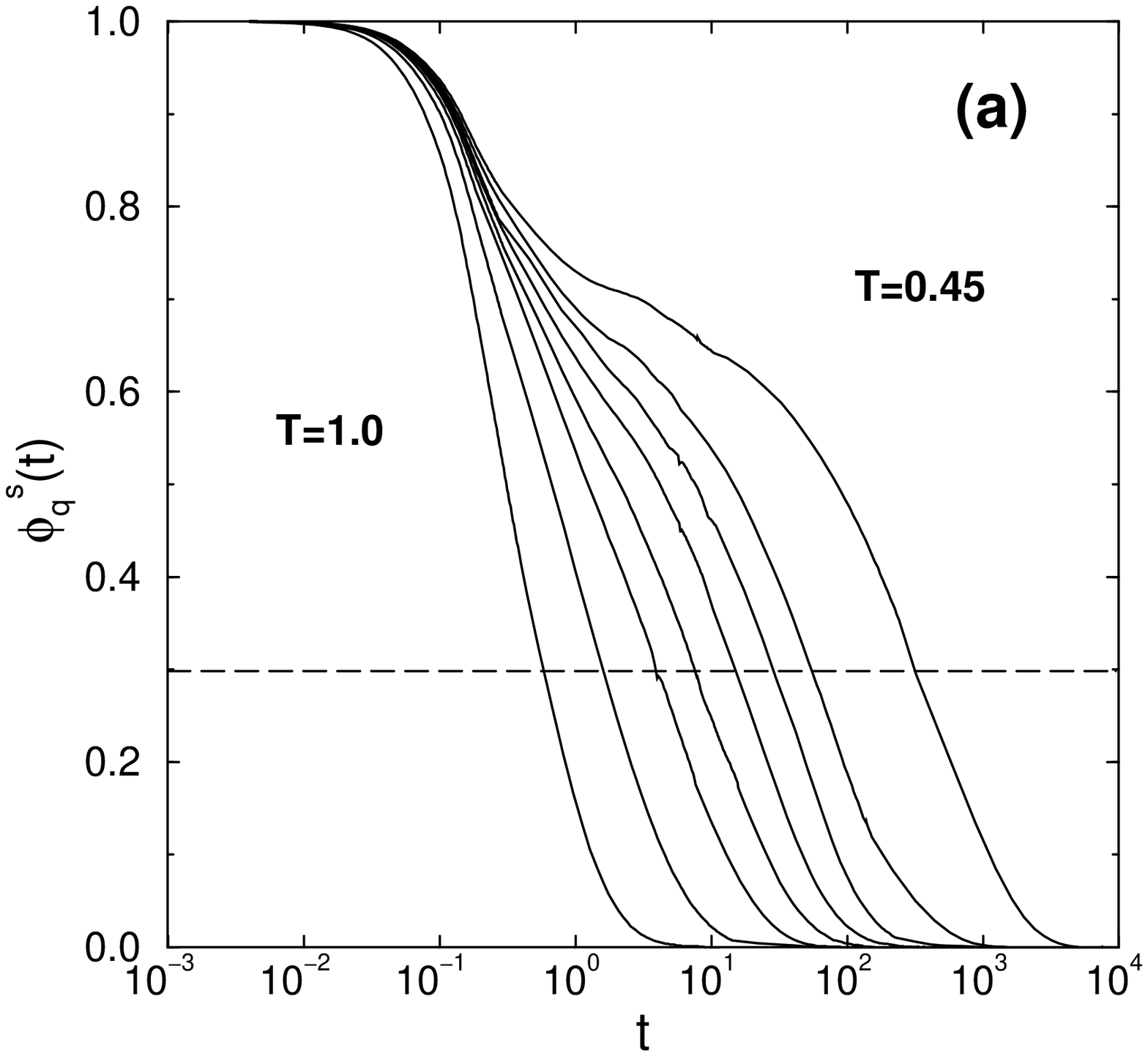}(a)
\includegraphics[width=110mm,height=90mm]{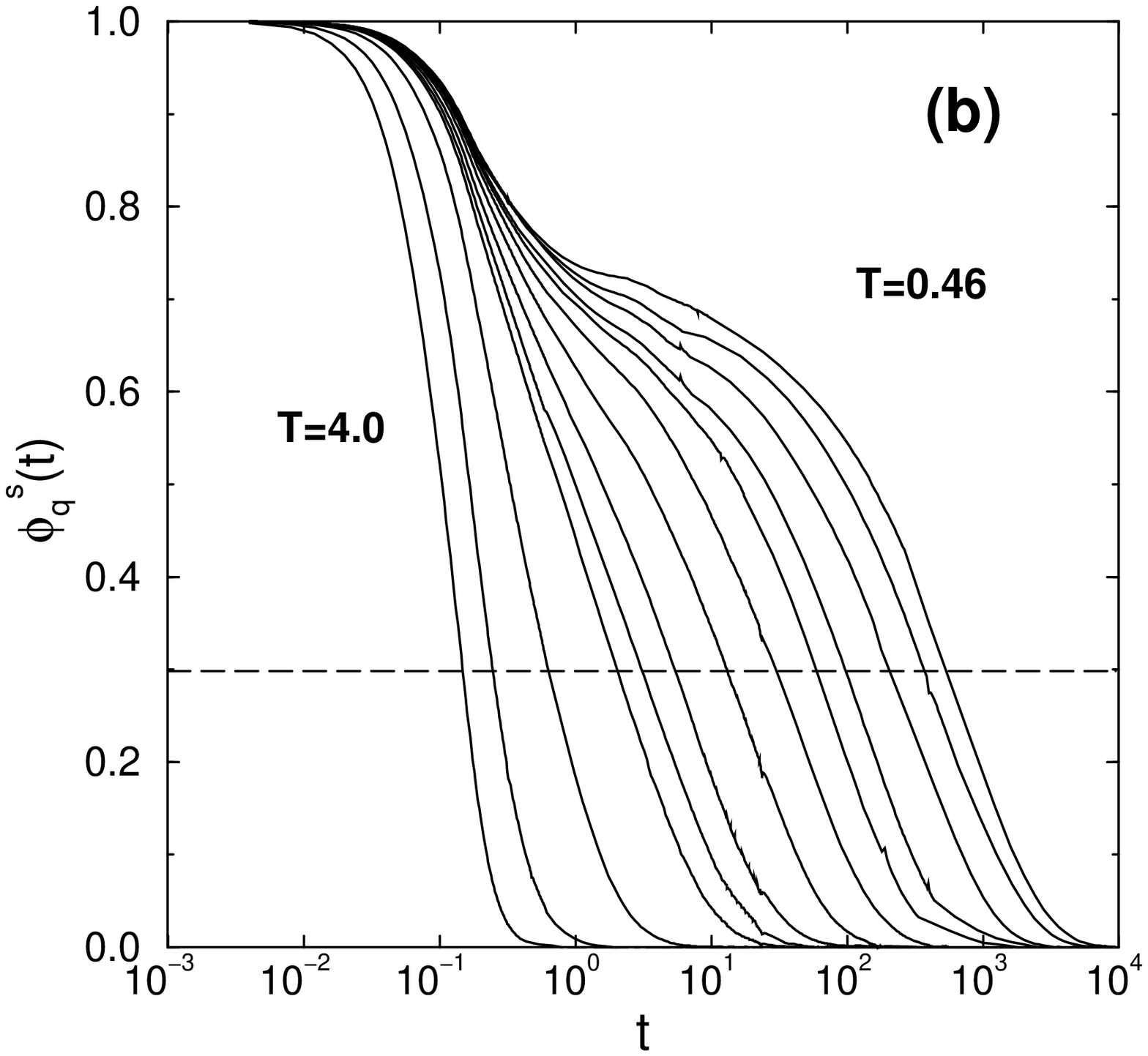}(b)
\end{figure}
\begin{figure}
\includegraphics[width=110mm,height=90mm]{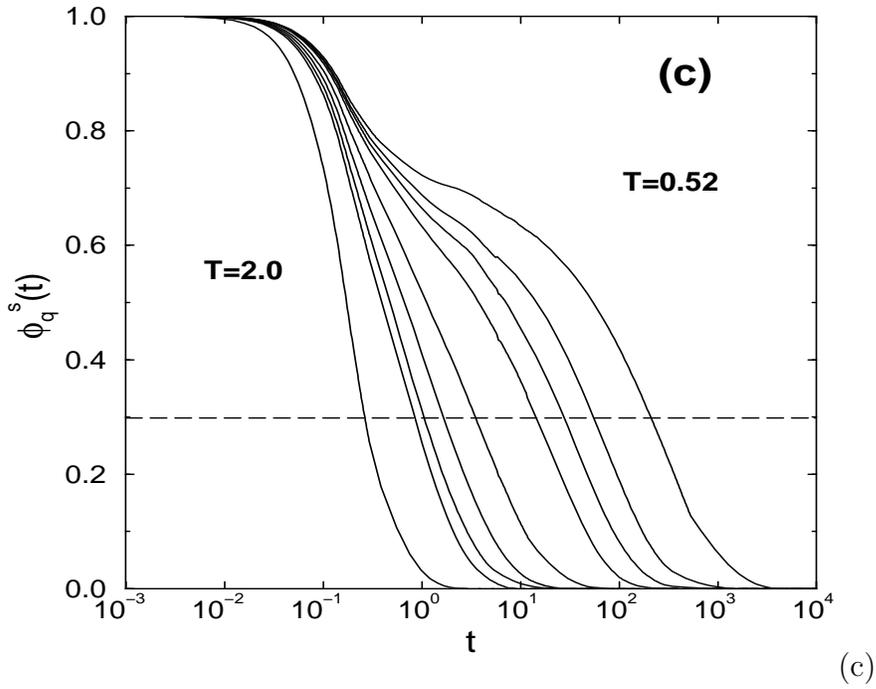}(c)
\caption{Intermediate dynamic structure factors at the first maximum of 
the static structure factor ($q=6.9$) \cite{supermct} measured along the isobars
$p=0.5$ (a), $p=1.0$ (b) and $p=2.0$ (c). The broken line shows
the value, which we used to define the $\alpha$-relaxation time scale.
From right to left, the temperatures decrease, as specified
in Table~\ref{tab1}.}
\label{fig1}
\end{figure}

\begin{figure}[h]
\centering
\includegraphics[width=110mm,height=90mm]{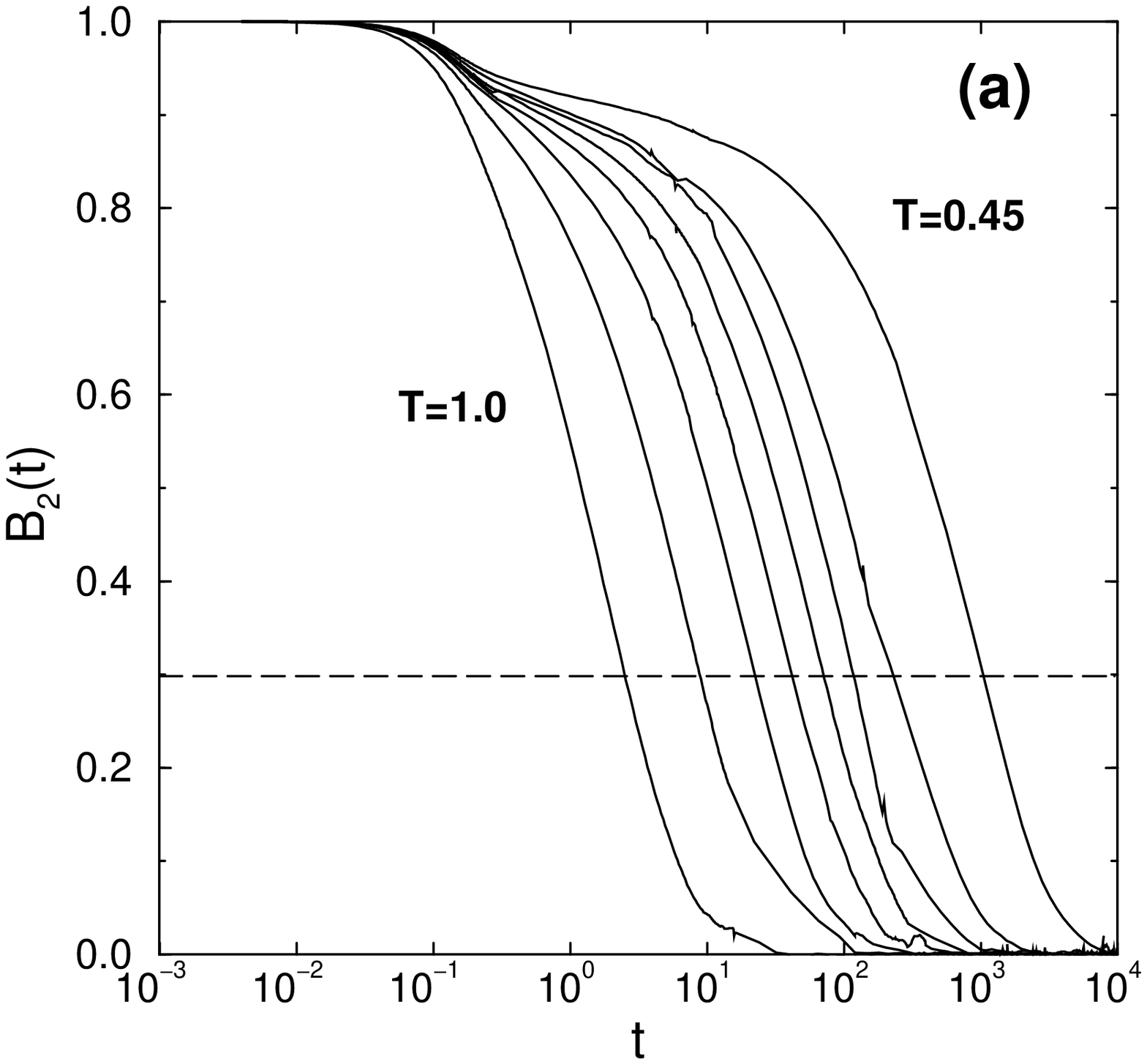}(a)
\includegraphics[width=110mm,height=90mm]{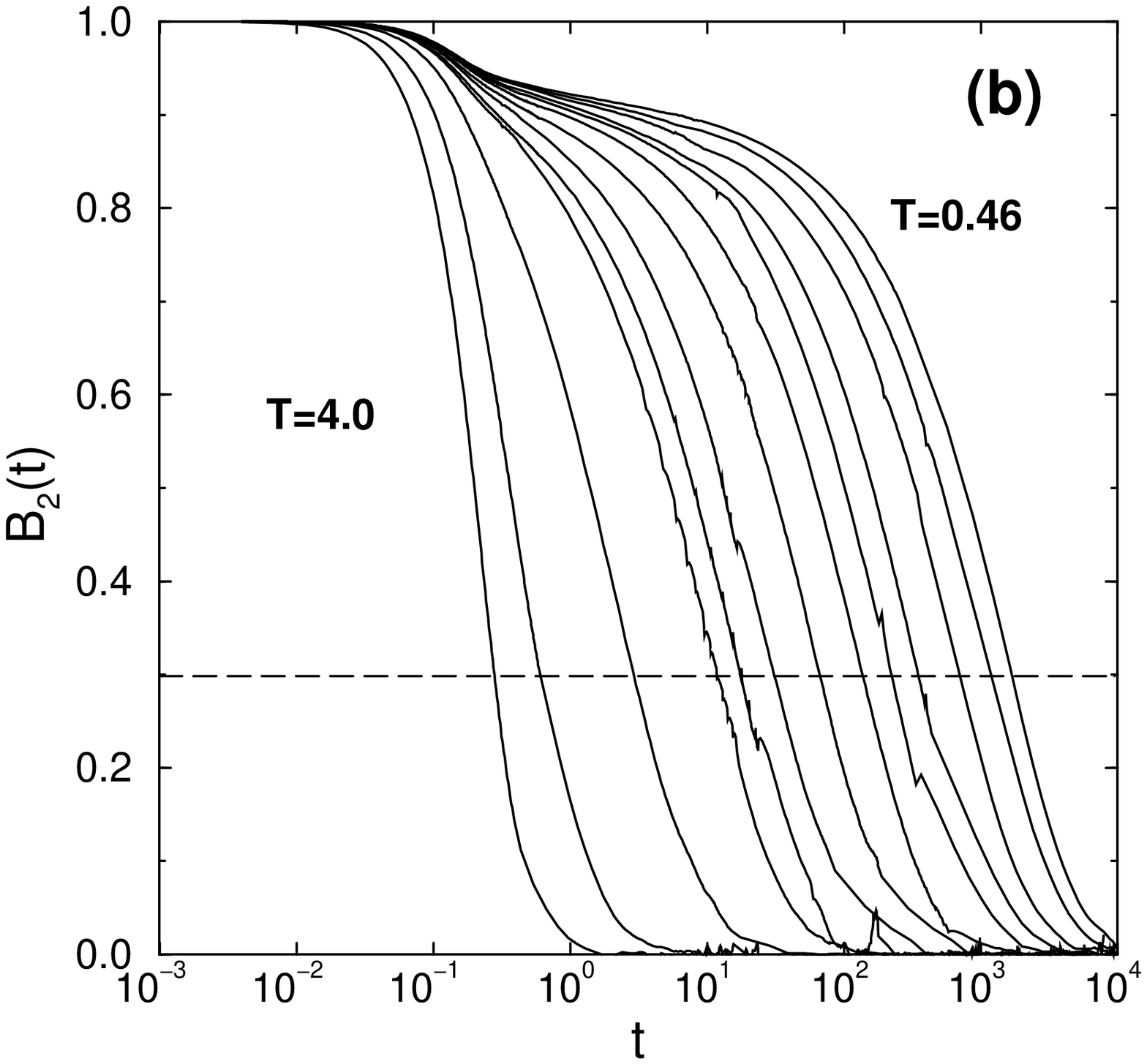}(b)
\end{figure} 
\begin{figure}
\includegraphics[width=110mm,height=90mm]{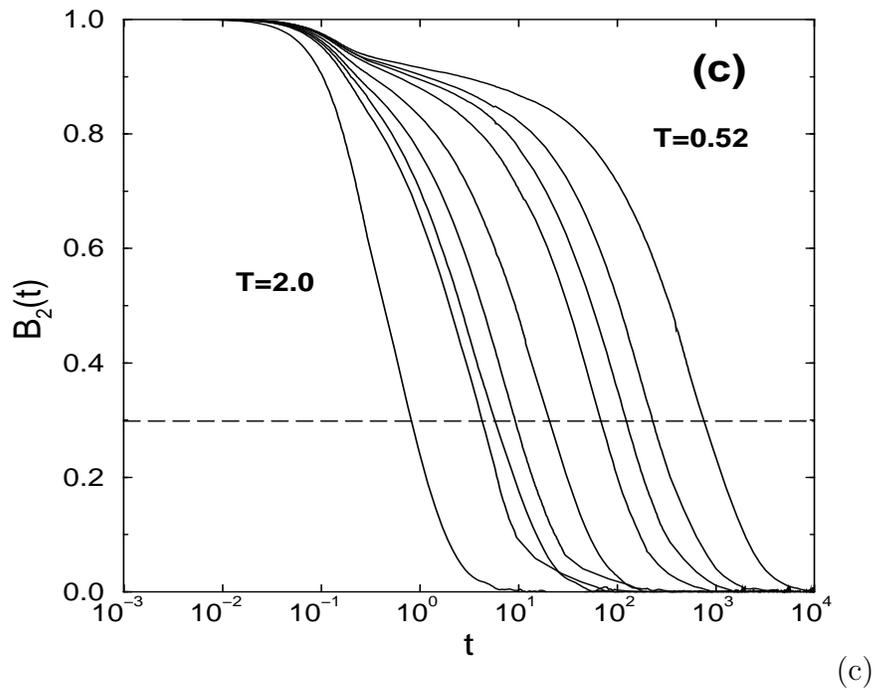}(c)
\caption{Dynamic correlation functions of the orientation of the bond vectors 
[second Legendre polynomial, see Eq.~(\ref{bn})]. Different pressures
are shown: $p=0.5$ (a), $p=1.0$ (b) and $p=2.0$ (c). From
right to left, the temperatures decrease, as specified in
Table~\ref{tab1}.}
\label{fig2}
\end{figure}

\begin{figure}[h]
\centering
\includegraphics[width=110mm,height=90mm]{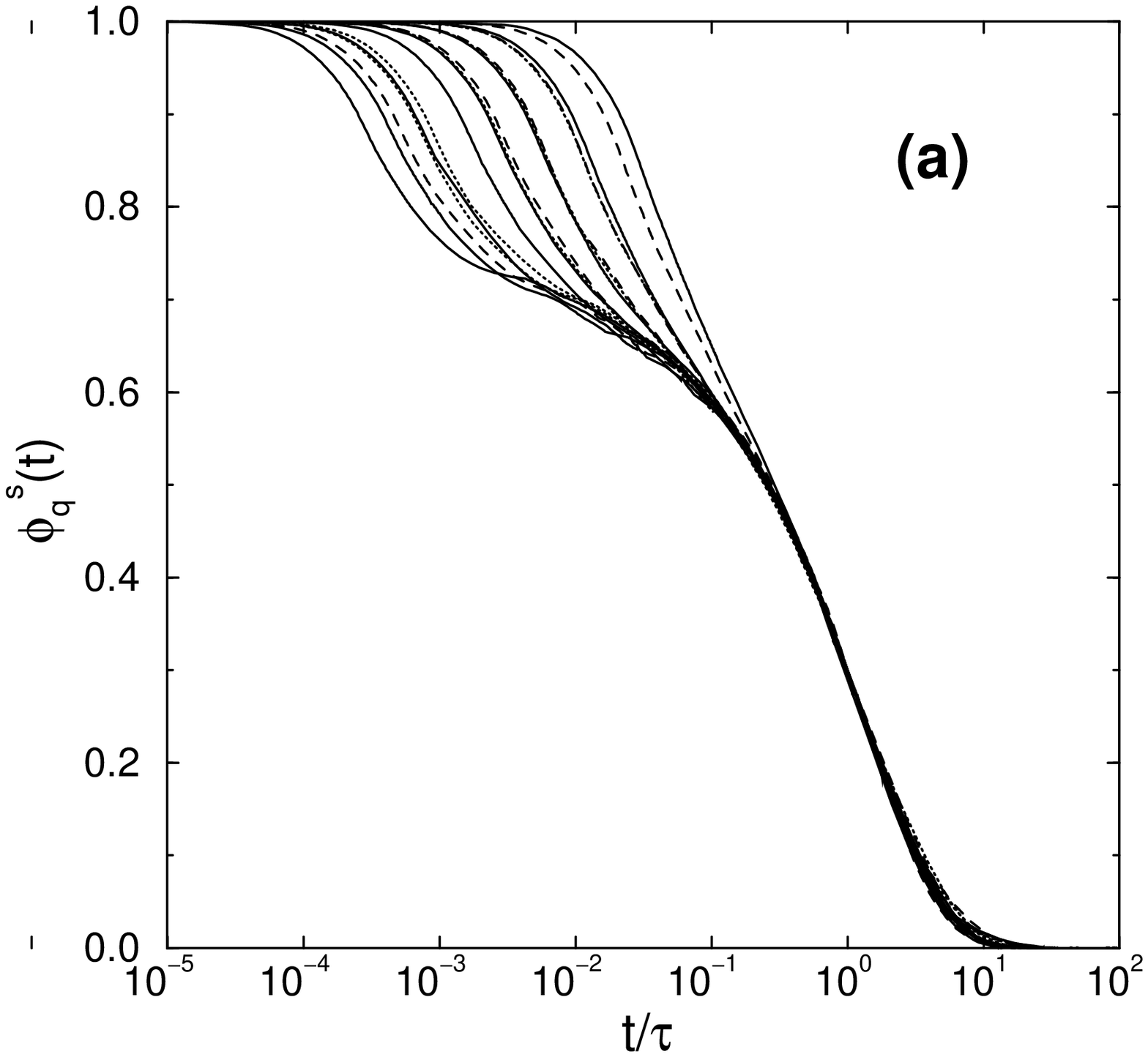}(a)
\includegraphics[width=110mm,height=90mm]{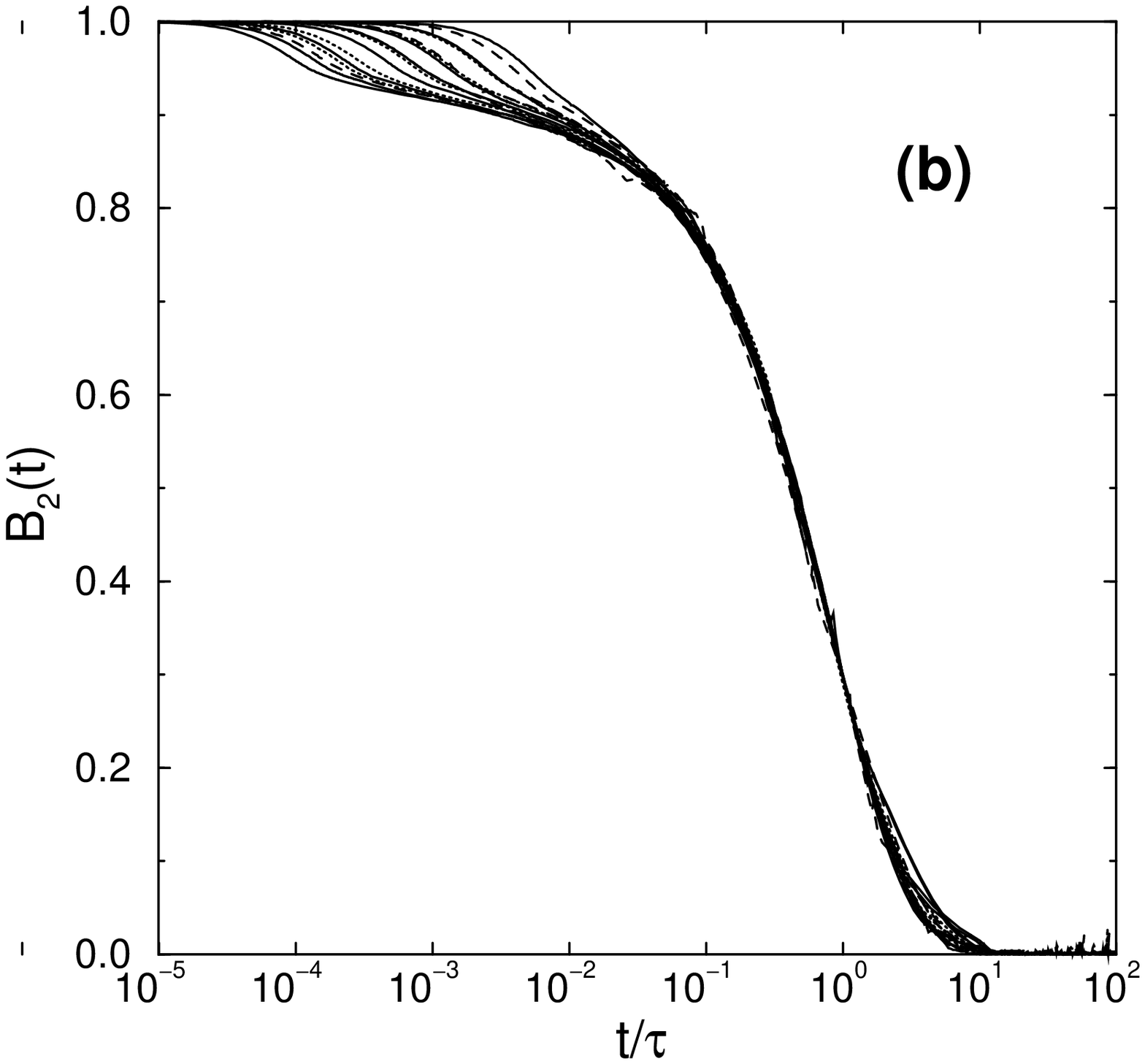}(b)
\end{figure} 
\begin{figure}
\includegraphics[width=110mm,height=90mm]{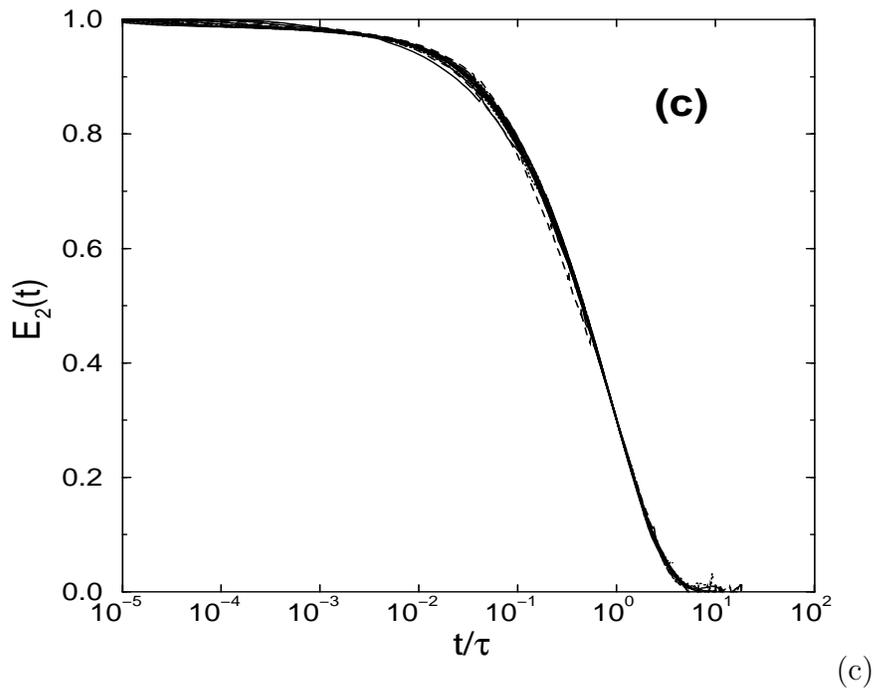}(c)
\caption{Figure.~3a is a compilation of the results from
Figs.~1a-c, but with times scaled by the $\alpha$-relaxation
time scale for selected temperatures close to the critical temperature at the respective
pressure: $T=0.45,0.48,0.5,0.52,0.55$ for $p=0.5$ (dashed lines), $T=0.46,0.47,0.48,0.5,0.52$ for $p=1$
(solid lines) and $T=0.52,0.55,0.57,0.6$ for $p=2$ (dotted lines). In the 
$\alpha$-regime the curves for different temperatures 
and pressures all collapse on a single master curve, 
demonstrating time-temperature-pressure superposition. 
Figures~3b and 3c show the same behavior for the 
orientational correlation functions $B_2(t)$ and $E_2(t)$ 
(second Legendre polynomial) of the bonds and the end-to-end
vector. Note that the plateau for $E_2(t)$ is so close to
1 that the first step cannot be seen on the scale of the
figure.}
\label{fig3}
\end{figure}

\begin{figure}
\centering
\includegraphics[width=110mm,height=90mm]{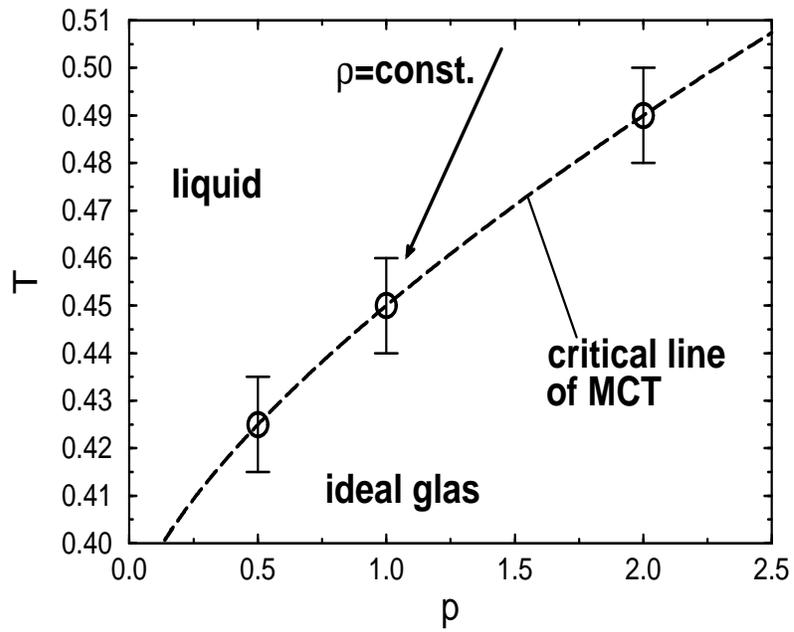}
\caption{Mode Coupling critical temperatures at different 
pressures. The critical temperatures represent
averages which are derived by fitting Eq.~(\ref{tgamma}) to
all relaxation times shown in Fig.~\ref{tau} . The broken 
line is an illustration of the critical line of MCT (guide
to the eye only), while the arrow symbolises
a thermodynamic path at constant density.}
\label{tcp}
\end{figure}

\begin{figure}
\centering
\includegraphics[width=110mm,height=90mm]{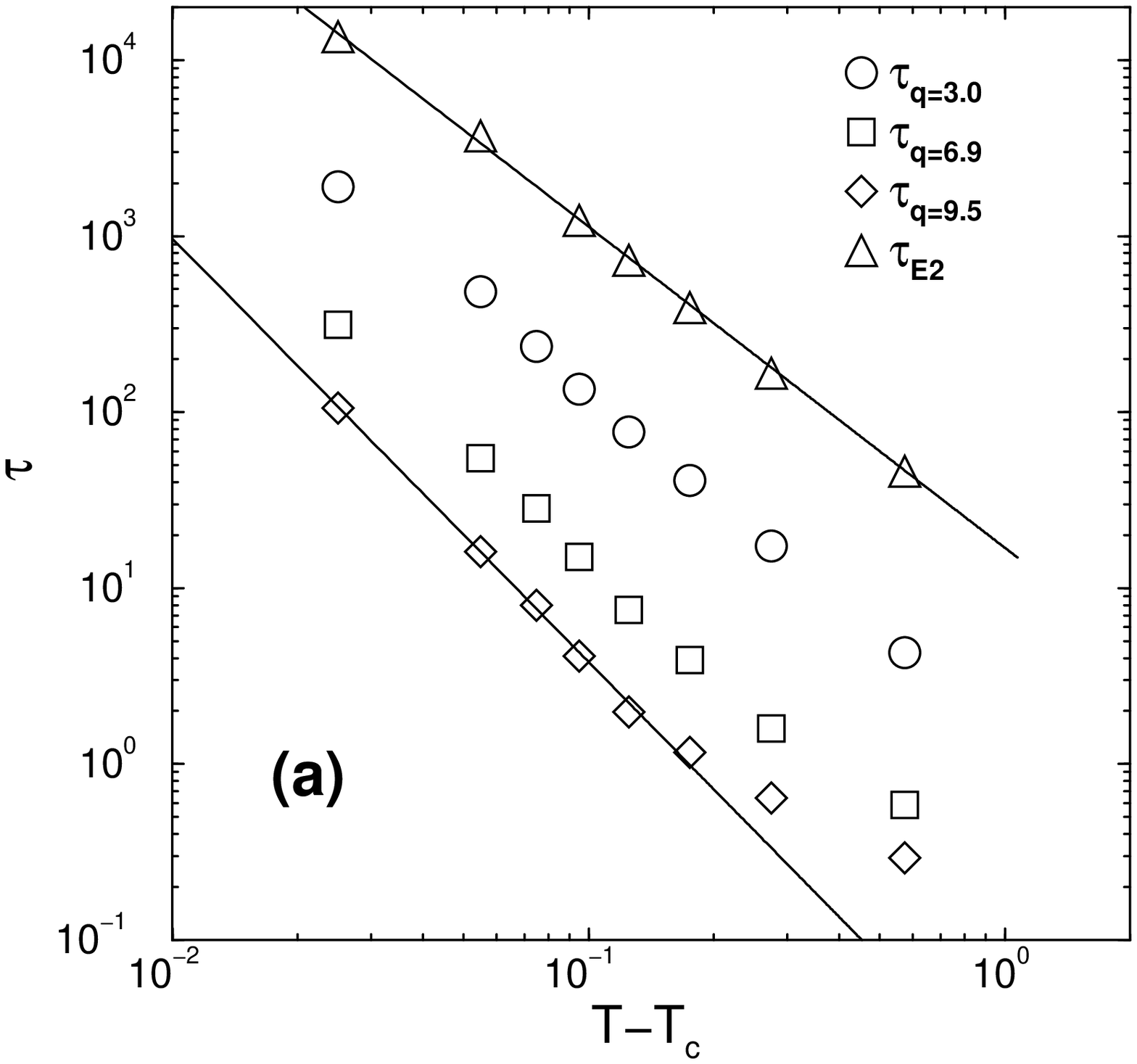}(a)
\includegraphics[width=110mm,height=90mm]{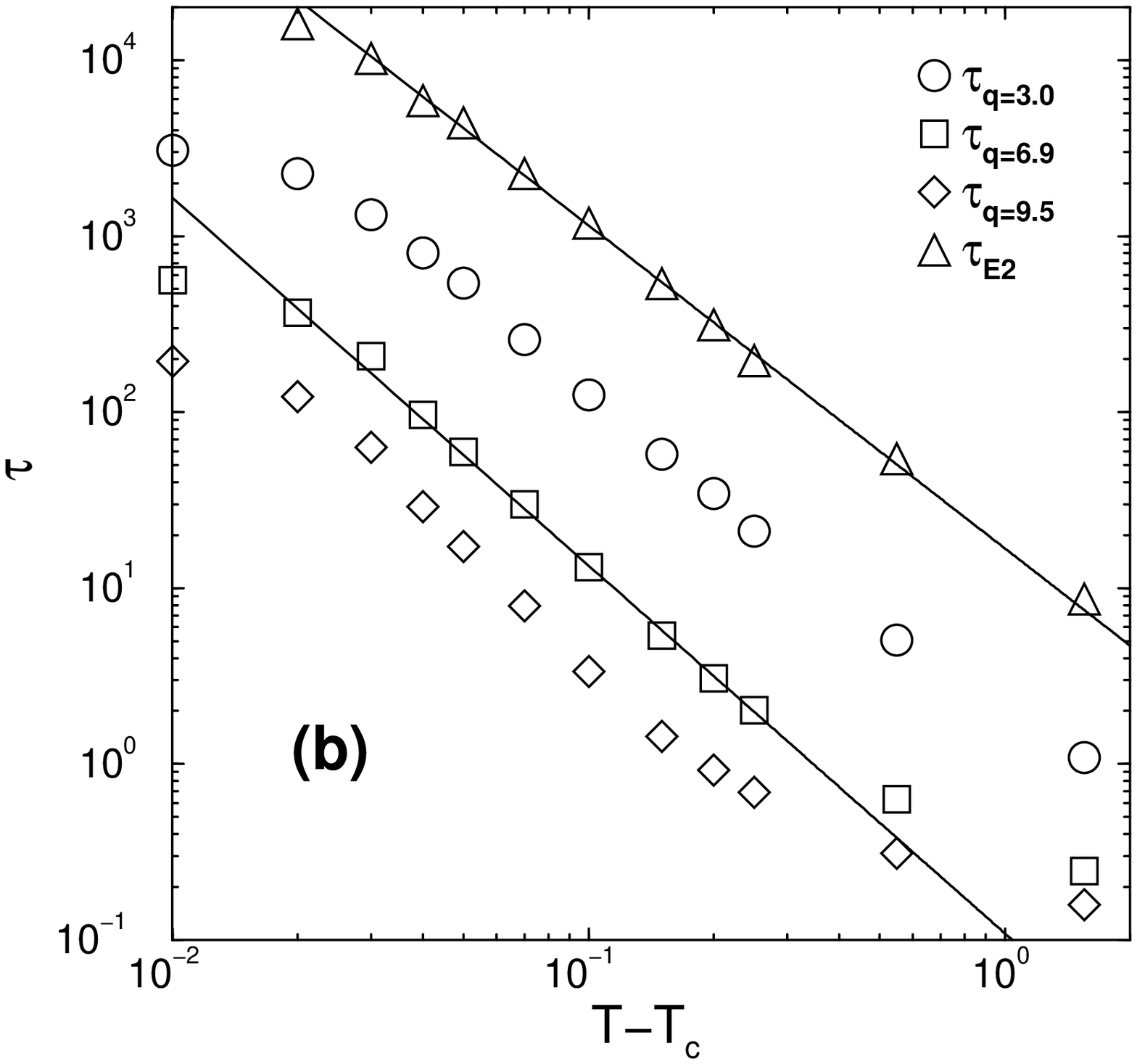}(b)
\end{figure} 
\begin{figure}
\includegraphics[width=110mm,height=90mm]{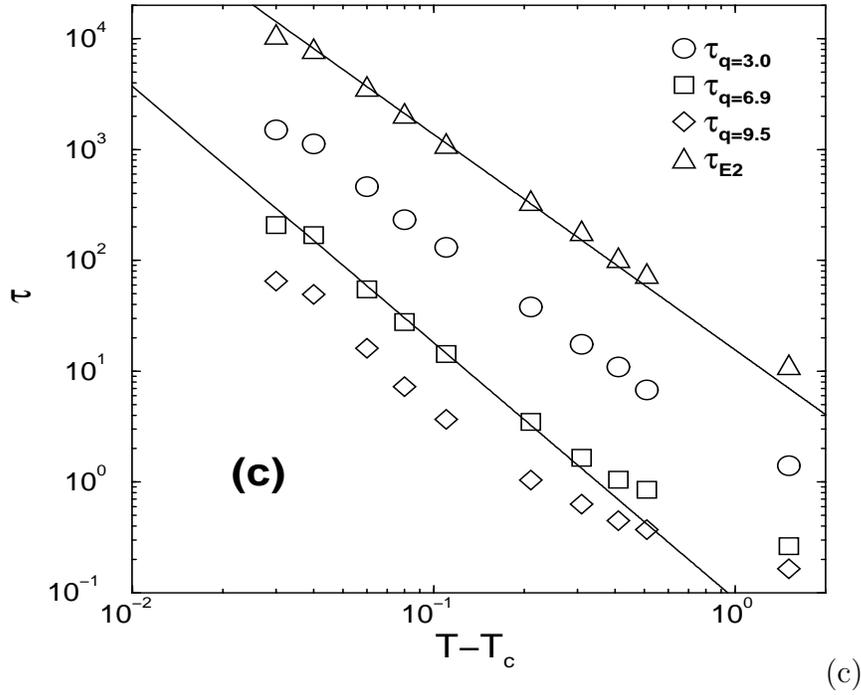}(c)
\caption{Temperature behaviour of different relaxation times, measured along
the isobars $p=0.5$ (a), $p=1.0$ (b) and $p=2.0$ (c). In the plots 
$\tau_q$ and $\tau_{E_2}$ are the $\alpha$-relaxation times of the incoherent dynamic structure factor
at different wave numbers and the dynamic orientational correlation of the end-to-end vector
(second Legendre polynomial), respectively. The values of
$T_c$ are taken from Eq.~(\ref{tcres}). The solid lines
are power-law fits, including the largest possible number
of temperatures. For $p=1$, the fit for $q=6.9$ uses $\gamma=2.09$, i.e.,
the $\gamma$-value resulting from an analysis of the 
$\beta$-relaxation \cite{supermct}.}
\label{tau}
\end{figure}

\begin{figure}[t]
\centering
\includegraphics[width=110mm,height=90mm]{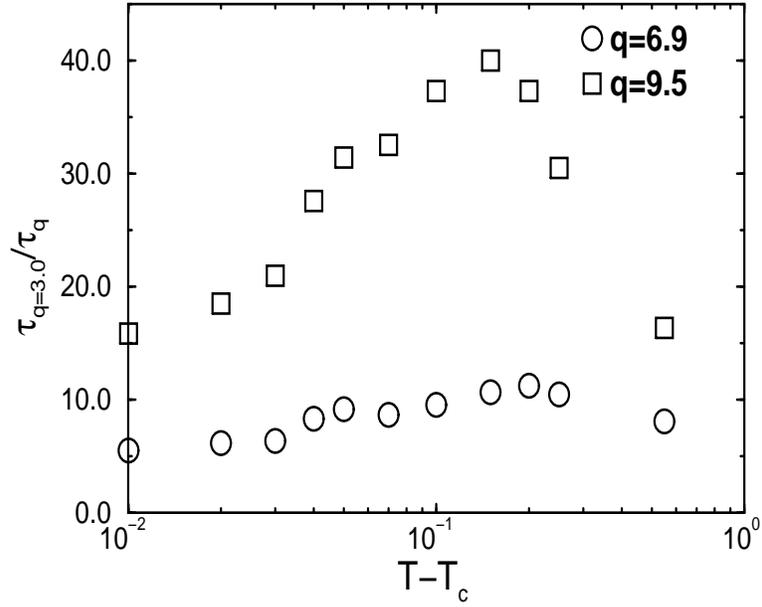}
\caption{Ratio of different $\alpha$-relaxation times as measured along the
isobar $p=1.0$. As can be seen, even close to the critical temperature ($T_c=0.45$)
the ratio changes by almost a factor of two.}
\label{ratio}
\end{figure}

\begin{figure}[h]
\centering
\includegraphics[width=110mm,height=90mm]{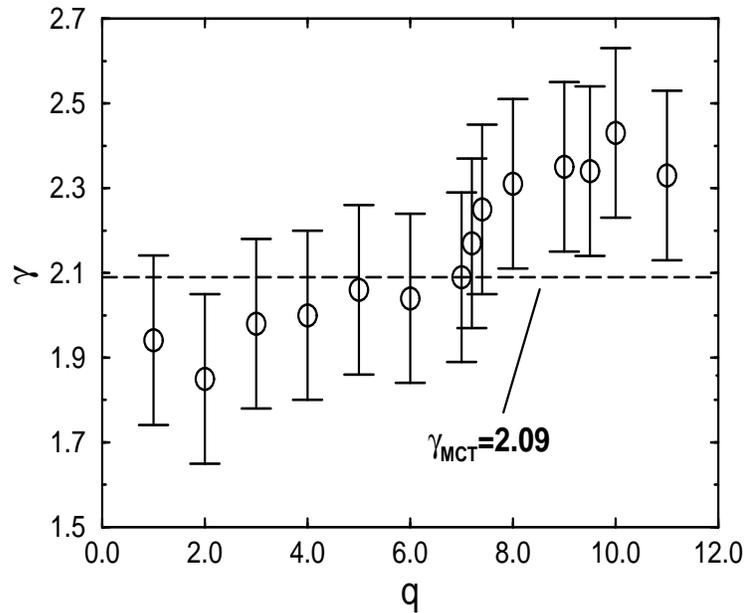}
\caption{Variation of $\gamma$ ($p=1.0$) with the magnitude
$q$ of the wave-vector when fitting the $\alpha$-relaxation
time of $\phi_q^\mathrm{s}(t)$ by Eq.~(\ref{tgamma}) while keeping
the critical temperature fixed ($T_c=0.45$).} 
\label{gammaq}
\end{figure}

\begin{figure}[h]
\centering
\includegraphics[width=110mm,height=90mm]{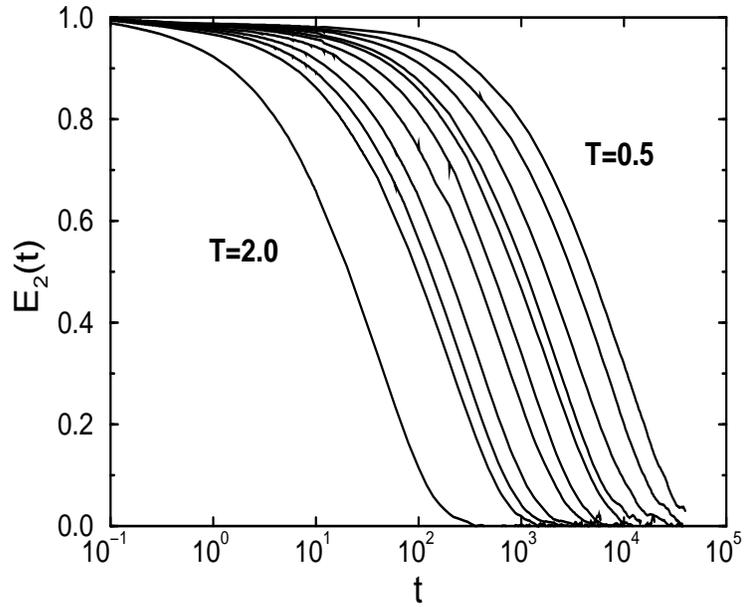}
\caption{Dynamic correlation of the end-to-end vector orientation (second Legendre polynomial),
as measured at constant density along the thermodynamic path $\rho=1.042$.
Temperature decreases from right to left, as specified in
Table~\ref{tab1}.}
\label{fig7}
\end{figure}

\begin{figure}[h]
\centering
\includegraphics[width=110mm,height=90mm]{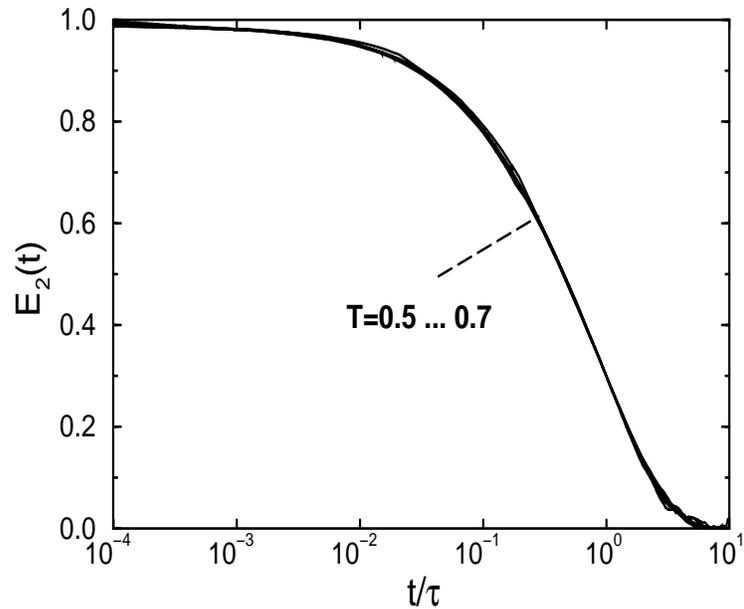}
\caption{$\alpha$-scaling plot of the end-to-end vector
correlation function for temperatures ranging from $T=0.5$
to $T=0.7$ (see Table~\ref{tab1} for details). The 
simulation data are for the same isochor as in Fig.~\ref{fig7},
and the relaxation time was determined by Eq.~(\ref{taudef}).}
\label{fig8}
\end{figure}

\begin{figure}[h]
\centering
\includegraphics[width=110mm,height=90mm]{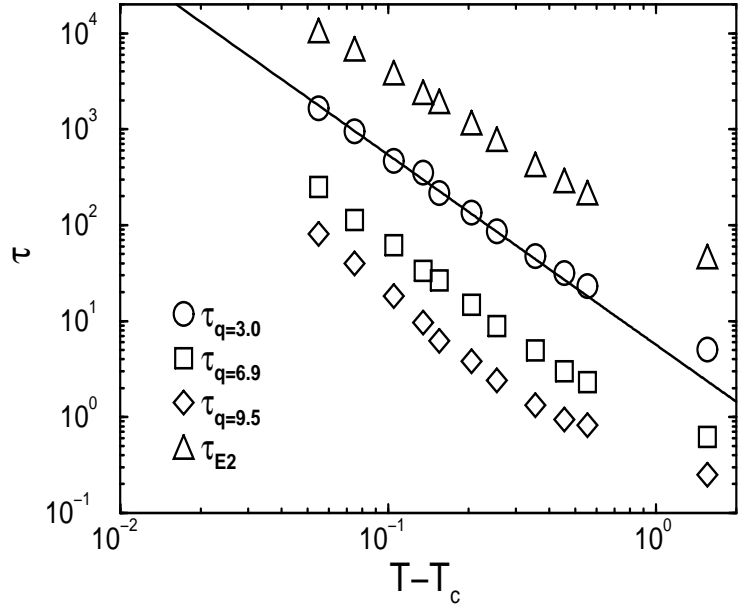}
\caption{Temperature behaviour of different $\alpha$-relaxation times, as measured 
in the NVT-ensemble. The solid line represents a fit with
Eq.~(\ref{tgamma}).}
\label{fig9}
\end{figure}

\begin{figure}[h]
\centering
\includegraphics[width=110mm,height=90mm]{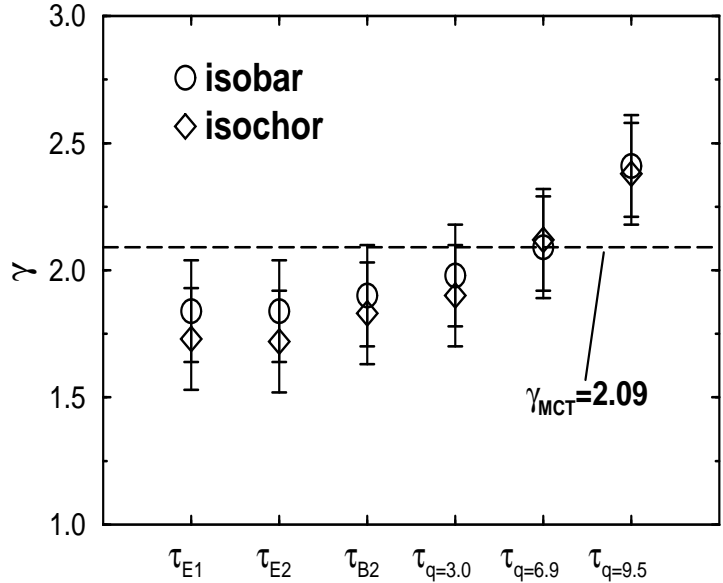}
\caption{Values of $\gamma$, determined from the
temperature dependence of various correlation times, for
two different thermodynamic paths, which yield the same
critical temperature. On the abscissa, the relaxation
times are quoted, from which $\gamma$ was determined.
For both the isobaric and the isochor path the error margins
are about $10 \%$, which is rather large, since $\gamma$ is 
very sensitive to a variation of the critical temperature. 
Within the error bars, $\gamma$ does not depend on the 
thermodynamic path.} 
\label{fig10}
\end{figure}

\end{document}